\documentclass[preprint,prd,noshowpacs,nofootinbib]{revtex4}

\usepackage{color} 
\usepackage{amssymb}
\usepackage{amsmath}
\usepackage{graphicx} 
\usepackage{float}
\usepackage{braket}     
\usepackage{pdfpages}
\usepackage{epstopdf}
\usepackage[utf8]{inputenc}

\begin{document}

\title{Quantum gravity, the cosmological constant, and parity transformation}

\author{Michael Bishop}
\email{mibishop@mail.fresnostate.edu}
\affiliation{Mathematics Department, California State University Fresno, Fresno, CA 93740}

\author{Peter Martin}
\email{kotor2@mail.fresnostate.edu}
\affiliation{Physics Department, California State University Fresno, Fresno, CA 93740}

\author{Douglas Singleton}
\email{dougs@mail.fresnostate.edu,  }
\affiliation{Physics Department, California State University Fresno, Fresno, CA 93740 \\}
\affiliation{Kavli Institute for Theoretical Physics, University of California Santa Barbara, Santa Barbara, CA 93106, USA}

\date{\today}

\begin{abstract}
One of the leading issues in quantum field theory and cosmology is the mismatch between the observed and calculated values for the cosmological constant in Einstein’s field equations of up to 120 orders of magnitude. In this paper, we discuss new methods to potentially bridge this chasm using the generalized uncertainty principle (GUP). We find that if quantum gravity GUP models are the solution to this puzzle, then it may require the gravitationally modified position operator undergo a parity transformation at high energies. 
\end{abstract}

\maketitle

\section{Cosmological Constant Puzzle}

The cosmological constant problem is the naive mismatch by $\sim$ 120 orders of magnitude between the observed and a simple theoretical value for the cosmological constant in Einstein's field equations; we will clarify in what sense this 120 order of magnitude mismatch is misleading/naive below.  Quantum gravity has long been advertised as a potential solution to various puzzles like this cosmological constant problem.   In this paper, we lay out the reasoning that led us to the conclusion that if GUP models are to resolve the cosmological constant problem, they would require the gravitationally modified position operator to undergo a parity transformation at high energy scales or short distance scales.

GUP models modify the Heisenberg uncertainty principle to obtain a minimum length scale; that is, a positive lower bound on the uncertainty in position.  
GUP models are not elegant, top-down theories of quantum gravity like string theory \cite{polchinski} or loop quantum gravity \cite{rovelli}, but they have the advantage of being easy to work with and provide a phenomenological window into quantum gravity. 
We will begin by briefly reviewing the salient features of the cosmological constant problem; a more complete discussion can be found in the excellent review by Weinberg \cite{weinberg}. We will adopt the notation and units of \cite{weinberg}. 

The cosmological constant, $\lambda$, in Einstein's field equations ($G_{\mu \nu} = - 8 \pi G T_{\mu \nu} + \lambda g_{\mu \nu}$) is equivalent to space-time having a constant energy density $\rho_{vac} = \frac{\lambda}{8 \pi G}$, and is representative of the expansion of space. 
The approach for the calculation of $\lambda$ in QFT is to add up all the energies of the zero modes (vacuum modes) of quantum fields. The vacuum modes of quantum fields are given by $\frac{1}{2} \hbar \omega _p = \frac{1}{2} E_p = \frac{1}{2} \sqrt{{\vec p}^2 +m^2}$. Integrating over all possible momenta up to some cut-off, $p_c$, yields the vacuum energy density
\begin{equation}
    \label{rho-vac}
    \rho_{vac} = \int \frac{1}{2} \sqrt{{\vec p}^2 +m^2} \frac{d^3p}{(2\pi)^3}  = \frac{1}{2} \int _0 ^{p_c} \frac{4 \pi}{(2 \pi)^3} dp ~ p^2 \sqrt{p^2 +m^2} \approx \frac{p_c ^4}{16 \pi^2} ~.
\end{equation}
Here, $p=|{\vec p}|$ is the magnitude of the momentum. We use $p$ and $|{\vec p}|$ interchangeably throughout this work.

The integral in \eqref{rho-vac} is divergent, so it needs to be cut off at some momentum/energy scale, $p_c$, which is usually taken to be the Planck scale, {\it i.e.} $p_c \sim (8 \pi G)^{-1/2}$. Inserting this value of $p_c$ into \eqref{rho-vac} gives $\rho _{vac} \approx 2 \times 10^{71}$ GeV$^4$. The observed vacuum energy density \cite{pdg} is about $\rho^{obs} _{vac} \approx 10^{-47}$ GeV$^{4}$, which is a difference of 118 orders of magnitude - a terrible disagreement between theory and observation. As mentioned in the opening paragraph this approximately 120 orders of magnitude mismatch is somewhat naive. First, the hard cut-off of $p_c$ in \eqref{rho-vac}, leading to a quartic dependence on $p_c$, violates relativistic invariance which requires $\rho = - p$ {\it i.e.} energy density equals the negative pressure. This was pointed in \cite{akhmedov,koksma}, and these two works also show that if one uses dimensional regularization of integrals like \eqref{rho-vac}, one can restore relativistic invariance, and the energy density is no longer quartically dependent on cut-off scale, but has a logarithmic dependence. Finally, in \cite{koksma} it was shown that by using dimensional regularization of integrals like those in \eqref{rho-vac} and using known Standard Model fields (specifically the top quark, $W^{\pm}, Z^0$ bosons and Higgs boson with an estimated mass of 150 GeV) gave a disagreement between $\rho^{obs} _{vac} \approx 10^{-47}$ GeV$^{4}$ and the theoretical value of ``only" 56 orders of magnitude. This is still terrible, but essentially cuts in half the order of magnitude disagreement, from $\sim$ 120 to $\sim$ 56. This estimate of \cite{koksma} is more realistic since the only scale probed experimentally is the electroweak scale of $\sim $ 100 GeV. Using the Planck scale gives an overestimate of the disagreement. Nevertheless using either the electroweak scale or the Planck scale gives a huge discrepancy between the theoretical and observed value of the vacuum energy density.

One of the standard ideas for addressing the cosmological constant problem is via {\it unbroken supersymmetry} \footnote{In unbroken supersymmetry particles and their superpartner particles have the same mass.} (SUSY) models \cite{weinberg,tasi}. SUSY models have equal numbers of bosonic and fermionic fields/degrees of freedom. Since bosonic and fermionic vacuum modes have opposite signs, their vacuum energies cancel one another exactly in unbroken SUSY and one would have a natural explanation for a cosmological constant that is exactly zero. However, one does not want an exactly zero cosmological constant and SUSY {\it is} broken at least up to some high energy scale  greater than $\Lambda _{SUSY} > 10^3$ GeV. But if one used  the lower limit of $\Lambda _{SUSY}=10^3$ GeV as the cut-off $p_c$, then the disagreement between theory and experiment is $\sim$ 60 orders of magnitude.  Therefore, simple SUSY models do not provide an answer to the cosmological constant puzzle. 

However, despite SUSY (either broken or unbroken) not providing a solution to the cosmological constant puzzle, it nevertheless has the feature that one obtains a small/zero cosmological constant by a cancellation of large positive and negative contributions. In the GUP approach to the cosmological constant problem presented here, we find that we can similarly get a small cosmological constant by having a modified position operator which flips parity at some large energy/momentum scale. This changing of the parity of the modified position operator leads to a cancellation of large positive and large negative contributions to the cosmological constant. Such violations of parity in the gravitational interaction were studied theoretically in \cite{leitner} and recently experimental bounds have been placed on such parity violations in gravity \cite{zhang-2023}.

\section{Quantum Gravity via Generalized Uncertainty Principle}

In this section, we  will lay out the GUP approach to quantum gravity \cite{vene,gross,amati2,amati,maggiore,garay,KMM,scardigli,adler-1999,adler-2001}. GUP models are phenomenological methods to quantize gravity which modify the canonical position and momentum operators and by extension their commutator.  This leads to a modified Heisenberg uncertainty principle which gives another avenue to analyze how quantum gravity works at short distances and high energies.   

Inspired by \cite{KMM}, we will only modify the position operator and keep the standard momentum operator: 
\begin{equation}
    \label{mod-op}
    {\hat X}_i = i \hbar f(|{\vec p}|) \frac{\partial}{\partial p_i} ~~~{\rm and}~~~ {\hat p}_j = p_j; 
\end{equation} 
the capitalized $\hat{X}_i$ indicates modification.  These operators have the modified commutator
\begin{equation}
    \label{mod-com}
    [{\hat X}_i, {\hat p}_j] = i  \hbar\delta _{ij} f(|{\vec p}|).
\end{equation}
In \cite{KMM}, $f(|{\vec p}|) = 1+\beta |{\vec p}|^2$ where $\beta$ is a phenomenological parameter which sets the scale for quantum gravity.  
Generally, $\beta$ is taken to be the Planck scale $\beta \sim \frac{l^2_{Pl}}{\hbar ^2}$. The modified commutator in \eqref{mod-com} implies an uncertainty relationship $\Delta X \Delta p \sim 1 + \beta (\Delta p)^2$ which leads to  
a minimum length of $\Delta {\hat X_j} = \hbar \sqrt{\beta}$ at $\Delta {\hat p_j} = \frac{1}{\sqrt{\beta}}$.
Additionally, in order for position and momentum operators to be symmetric, {\it i.e.} $(\langle \psi | {\hat p}_i ) | \phi \rangle = \langle \psi | ({\hat p}_i  | \phi \rangle)$ and $(\langle \psi | {\hat X_i} ) | \phi \rangle = \langle \psi | ({\hat X_i}  | \phi \rangle)$, the scalar product of this model must have the form
    \begin{equation}
        \label{scalar}
        \langle \psi | \phi \rangle = \int _{- \infty} ^{\infty} \frac{d^3p}{f(|{\vec p}|)} \psi ^* (p) \phi (p).
    \end{equation}
The modification of the scalar product as given by \eqref{scalar} is for three dimensions; even in $n$ dimensions one still has the same modifying factor for the momentum integration, $\frac{d^np}{f(|{\vec p}|)}$. 

\section{Vacuum Energy Calculations with GUP}

With these modified operators and scalar products, one can modify the vacuum energy integration in \eqref{rho-vac} following \cite{chang}. In \cite{chang}, the authors calculate how the GUP  modifies Liouville's theorem and the phase space volume,  $d^3 x ~d^3 p$.
The modified phase space associated with the model from \eqref{mod-op} and \eqref{mod-com} is
\begin{equation}
    \label{phase-space-1}
    \frac{d^3 x ~ d^3 p}{( f(|{\vec p}|))^3}.
\end{equation}
Upon integrating out the volume, $\int d^n x \to V$, and after quantization, the phase space volume is given by 
\begin{equation}
    \label{phase-space-2}
    \frac{V ~ d^3 p}{(2 \pi \hbar)^3 (f(|{\vec p}|))^3} \to   \frac{V ~ d^3 p}{(2 \pi )^3 (f(|{\vec p}|))^3}.
\end{equation}
In the last step in \eqref{phase-space-2}, we set $\hbar \to 1$ to match the units of reference \cite{weinberg}. Using the modified phase space volume in \eqref{phase-space-2} for the calculation of $\rho_{vac}$ in \eqref{rho-vac} gives
\begin{equation}
    \label{rho-vac-1}
    \rho_{vac} = \int \frac{1}{2} \sqrt{{\vec p}^2 +m^2} \frac{d^3p}{(2\pi)^3 (f(|{\vec p}|))^3}.
\end{equation} This is beneficial because the factor of $(f(|{\vec p}|))^3$ makes the integrand integrable and makes the vacuum energy density finite without using a `by hand' cut-off.

Substituting $f(|{\vec p}|) = (1+ \beta |{\vec p}|^2)$ from \cite{KMM, chang} into \eqref{rho-vac-1}, one finds that the GUP vacuum energy density is finite with $\rho_{vac} \propto \beta ^{-2}$. However, $\beta$ is typically taken to be of the Planck scale and this leads to the GUP vacuum energy density from \eqref{rho-vac-1} yielding the same value as the by hand cut-off value obtained from \eqref{rho-vac}. Unfortunately, nothing has been gained by exchanging the by hand cut-off for a GUP-inspired functional cut-off, as already noted in \cite{chang}. 
Moreover, the inner product used for the vacuum energy density in \eqref{rho-vac-1} does not preserve the symmetry of the modified position operator. Equations \eqref{phase-space-1}, \eqref{phase-space-2}, and \eqref{rho-vac-1} imply that integration over momentum should come with the factor $f(|{\vec p}|))^{-3}$. This violates the symmetry requirement for the position operator $(\langle \psi | \hat{x}_i ) | \phi \rangle = \langle \psi | (\hat{x}_i  | \phi \rangle)$. In order for the position operator to be symmetric, one needs the factor  $f(|{\vec p}|))^{-1}$ as in equation \eqref{scalar}. 

To resolve this conflict between  \eqref{scalar} and \eqref{phase-space-2}, we need to reconsider the spatial/volume calculation. In the transition from \eqref{phase-space-1} to \eqref{phase-space-2}, it is assumed that the spatial volume does not change, {\it i.e.} the spatial  volume is $\int d^n x = V$; but since this approach utilizes the GUP, one would imagine that the introduction of a minimal length should change the calculation of volumes. 

We pair a single factor of $f(|{\vec p}|)$ to go with the $d^3p$ integration, which would be consistent with the symmetry of the position operator required by \eqref{scalar}. 
The remaining two factors are paired with the volume integration to account for the introduction of a minimal length.
The phase space integration with the spatial volume in spherical coordinates becomes \cite{pete-2023}  
\begin{equation}
    \label{phase-space-4}
    \left( \frac{d^3 x}{(f(|{\vec p}|))^2} \right) \left( \frac{d^3 p}{f(|{\vec p}|)} \right) 
    =    \left( \frac{r^2 dr d\Omega}{(f(|{\vec p}|))^2} \right) \left( \frac{d^3 p}{f(|{\vec p}|)} \right)~.
\end{equation}
The finite length factor in \eqref{phase-space-4} is $r^2$, so that the spatial integration that takes into account the GUP minimal length is $\frac{r^2}{f(|{\vec p}|))^2}$.  
Now, the modified momentum integration in \eqref{phase-space-4} agrees with the requirement of symmetry of the position and momentum operators as given in \eqref{scalar}. The corrected GUP-modified vacuum energy is then
\begin{equation}
    \label{rho-vac-2}
    \rho_{vac} = \int \frac{1}{2} \sqrt{{\vec p}^2 +m^2} \frac{d^3p}{(2\pi)^3 f(|{\vec p}|)}  ~.
\end{equation}
Whether we use \eqref{rho-vac}, \eqref{rho-vac-1},  or \eqref{rho-vac-2} to calculated $\rho_{vac}$, the discrepancy between the calculated and observed vacuum energy density is still enormous. This discrepancy is due to the fact that $f(|{\vec p}|)$ is always positive. 

One possible resolution is to allow $f(|{\vec p}|)$ to become negative at large momentum. With this modification, the vacuum energy density integral has a negative contribution which can bring the calculated vacuum energy density closer to the observed value.  Generally, GUP models do not consider an $f(|{\vec p}|)$ which can be negative, because this results in a parity flip of the position operator, {\it i.e.} $({\hat X}) \to - ({\hat X})$, at large momentum. Admittedly having a parity flip like this is very unusual and may lead to difficulties coming up with a good physical interpretation. However, the weak interaction is known to violate parity, and additionally there are works that have examined parity violation in gravity. For example, reference \cite{leitner} examined parity violation in gravity, and recently reference \cite{zhang-2023} placed experimental bounds on parity violation and time-reversal symmetry violation in gravity using spin-gravity interactions. Further in \cite{lue} a model of parity violation is constructed with potential signatures appearing in the cosmic microwave background (CMB). As another example references \cite{calcagni1,calcagni2} attempt to solve the cosmological constant problem in a loop quantum gravity model with degenerate geometry, with an implied parity violation. Thus while there are certainly questions as to what a parity flip like $({\hat X}) \to - ({\hat X})$ means physically, there is work examining parity violation in gravity both theoretically and experimentally.

In the next section we investigate the possibility that quantum gravity may lead to a modified position operator which changes sign/violates parity at some high energy/momentum scale leading to a small cosmological constant consistent with observations.   

\section{GUP Cosmological constant and Parity Transformation}
 
In order to recover the standard position operator at low energy scales, the function $f(|{\vec p}|)$ which modifies the position operator  must satisfy $f(|{\vec p}|) \approx 1$ when $|{\vec p}| \ll p_M$, where $p_M$ designates the momentum where the parity flip occurs. 
As mentioned previously, one expects $f(|{\vec p}|) \approx 1$ up to, for example, the electroweak scale with $|{\vec p}| \approx 100$ GeV.  At this point the calculated $\rho_{vac}$ is already $\sim $56 orders of magnitude larger than the observed vacuum energy density. 

As discussed in the previous section, if $f(|{\vec p}|)$ is positive definite, then the integration from the electroweak scale upward will only make further positive contributions to $\rho_{vac}$, increasing the discrepancy between the calculated and observed vacuum energy density. 
The only way we can see to counter this large reserve of positive vacuum energy density is to have the vacuum density integrand in \eqref{rho-vac-2} become negative as the energy/momentum scale increases. 
This is conceptually similar to unbroken supersymmetry, where the positive contribution of bosonic zero modes balance the negative contribution of fermionic zero modes.

An example of a GUP model which satisfies the above requirement is 
\begin{equation}
    \label{x-cut}
    {\hat X}_i = i \hbar \left[1 - \left(\frac{|\vec{p}|}{p_M} \right)^2 \right]^{-1} \exp \left(\frac{2 |\vec{p}|^2}{p_N ^2}\right)\partial_{p_i} ~~~~;~~~~ {\hat P}_i = p_i~,
\end{equation}
where we have introduced a second momentum scale, $p_N$. We will see the need for this later, but $p_N$ should be of the order of $p_M$.

If we do an expansion to second order in $|{\vec p}|/p_M$ and , $|{\vec p}|/p_N$ we find that \eqref{x-cut} is equivalent to the GUP model of reference \cite{KMM} with  ${\hat X}_i \approx i \hbar (1 + \beta |\vec{p}|^2 ) \partial_{p_i}$ and ${\hat P}_i = p_i$, and $\beta$ depending on $p_M$ and $p_N$. The specific form of the modified position operator in \eqref{x-cut} is driven by two constraints: (i) we want the modified operators in \eqref{x-cut} to give a minimum length; (ii) we want the position operator to flip sign at some large momentum scale so that the vacuum energy density integral will have positive (at low momentum) and negative (at high momentum) contributions. This flipping of the sign of the position operator can be seen as a form of parity violation; which proposes that the gravitational interaction may violate parity, as is also the case for the weak interaction. 

Taking the modified operators from \eqref{x-cut} and using them to calculate the vacuum energy density \eqref{rho-vac-2}, taking into account that $f(|{\vec p}|) =  \left[1 - \left(\frac{|\vec{p}|}{p_M} \right)^2 \right]^{-1} \exp \left(\frac{2|\vec{p}|^2}{p_N ^2}\right)$, we get
\begin{eqnarray}
    \label{rho-vac-3}
    \rho_{vac} &=& \frac{1}{2} \int \sqrt{{\vec p}^2 +m^2} \left[1 - \left(\frac{|\vec{p}|}{p_M} \right)^2 \right] \exp \left(-\frac{2|\vec{p}|^2}{p_N ^2}\right) \frac{d^3p}{(2\pi)^3}  \nonumber \\
    &\approx& \frac{1}{4 \pi ^2} \int _0 ^\infty p^3 \left[1 - \left(\frac{p}{p_M} \right)^2 \right] \exp \left(-\frac{2p^2}{p_N ^2}\right) dp  \\
    &=& \frac{p_N ^4}{32 \pi ^2} \left( 1 - \frac{p_N ^2}{p_M ^2} \right) \nonumber ~.
\end{eqnarray}
From the first line to the second, we have written $|{\vec p}|$ simply as $p$, we have done the integration over the solid angle giving $4 \pi$, and we have assumed that $m^2 c^2$ is small compared to both $p^2 _M$ and $p^2 _N$ allowing us to use $\sqrt{p^2 + m^2} \approx p$. 

From the last expression in \eqref{rho-vac-3}, we can see a balancing between the positive contribution $p^4_N$ coming from the integration over low $p$, and the negative contribution $\frac{p^6_N}{p_M^2}$ coming from the integration over high $p$. This is reminiscent of the balancing of positive and negative contributions to the vacuum energy density in unbroken supersymmetric models. 

One can use \eqref{rho-vac-3} to ``solve" the cosmological constant problem. Setting the calculated vacuum energy density in \eqref{rho-vac-3} to the observed vacuum energy density, one can solve for $p_M$ and obtain
$$p_M = p_N \left( 1 - \frac{32 \pi^2 \rho _{vac} ^{obs}}{p_N ^4}\right)^{-1/2}.$$
If $p_N$ is at the Planck scale this implies $\frac{\rho _{vac}^{obs}}{p_N ^4} \sim 10^{-118}$, which in turn leads to $p_M \approx p_N$, and $p_M$ and $p_N$ are both of the Planck scale. 

Furthermore $p_{M}$ and $p_{N}$ do not necessarily need to be at the Planck scale to resolve the cosmological constant problem -- one just needs $\rho_{vac}^{obs} \ll p_N ^4$, which can be obtained even with $p_N$ and $p_M$ at a much lower scale than the Planck scale. One still has the fine tuning problem of why the two scales should deviate from the condition $p_M = p_N$ by such an incredibly small amount compared to either $p_M$ or $p_N$.
Nonetheless, this example illustrates how a negative contribution to the vacuum energy density, from high momentum, helps the GUP approach to the cosmological constant problem. 

One could ask if this model would be able to connect the present small value of $\rho _{vac} ^{obs}$ with a much larger value of the vacuum energy density required for an inflationary epoch in the very early Universe. Looking at the last line of \eqref{rho-vac-3} one can have a large $\rho _{vac}$ if, in the early Universe, the two momentum scales satisfy $p_M > p_N$, but not $p_M \approx p_N$. Then from this initial state the momentum scales would need to evolve toward $p_M \approx p_N$ to give the small observed vacuum energy density of the present Universe. 

\section{Summary and Conclusions}

In this paper we examined how GUP models might address the cosmological constant problem. For GUP models given by modified position and momentum operators of the form \eqref{mod-op} ${\hat X}_i = i \hbar f(|{\vec p}|) \frac{\partial}{\partial p_i}$ and ${\hat p}_j = p_j$, there were two variants of the associated modified vacuum energy density given in \eqref{rho-vac-1} and \eqref{rho-vac-2}. These two expressions differed by the number of factors of $\frac{1}{f(|{\vec p}|)}$. We argued for the validity of the expression in \eqref{rho-vac-2} over \eqref{rho-vac-1}, since only the former expression satisfied the requirement of symmetry of the modified position and momentum operators discussed around equation \eqref{scalar}.

However, regardless of whether one used \eqref{rho-vac-1} or \eqref{rho-vac-2} to calculate $\rho _{vac}$, one still has essentially the same problem as the by-hand cut-off of \eqref{rho-vac}: the vacuum energy density was proportional the momentum cut-off scale to the fourth power, $\rho _{vac} \sim p_c ^4$, which for $p_c$ near the Planck scale (or even the electroweak scale) made
$\rho _{vac}$ too large. With the GUP modified energy density of either \eqref{rho-vac-1} or \eqref{rho-vac-2} $\rho_{vac}$, while formally finite, would nevertheless be proportional to the inverse square of the functional cut-off parameter $\beta$ {\it i.e.} $\rho_{vac} \sim \beta ^{-2}$, and if the scale of $\beta$ were taken as the Planck scale, then one finds the same problem as using a by-hand cut-off of the formally infinite integral in \eqref{rho-vac} - the vacuum energy density from GUP models will have the same, large, order-of-magnitude disagreement compared to the observed vacuum energy density.  

We propose that the modified position operator changes sign at some momentum scale, $p_M$, as in \eqref{x-cut}. Then because of the link between the function $f(|{\vec p}|)$ from the modified position operator in  \eqref{mod-op}, and how it changes the vacuum energy density in \eqref{rho-vac-2}, one finds that the positive contribution to $\rho_{vac}$ from the integration below $p_M$ is balanced by negative contribution from above $p_M$
This balancing of large positive and negative contributions to the vacuum energy density is similar to unbroken SUSY, where a large, positive bosonic contribution is balanced by a large, negative fermionic contribution. Here, however this balancing of positive and negative contributions comes from the parity flip - the change in sign of the modified position operator at some momentum scale. \\

{\bf Acknowledgment:} DS is supported by a 2023-2024 KITP Fellows Award. This research was supported in part by the National Science Foundation under Grant No. NSF PHY-1748958. The work of MB and DS were supported through a Fresno State 2023-2024 RSCA grant.\\

\end{document}